\magnification=\magstep1
\hsize 32 pc
\vsize 42 pc
\baselineskip= 24 true pt
\def\cl {\centerline}
\def\vs {\vskip .4 true cm}
\cl {\bf An Understanding of The Dark Matter in The Universe And The }
\cl {\bf Variation of The Universal Gravitational Contant G With Time}
\vs
\cl {\bf D.N. Tripathy and Subodha Mishra}
\vs
\cl {Institute of Physics, Sachivalaya Marg, Bhubaneswar-751005, INDIA}
\vskip 0.5 true cm
\cl {\bf Abstract}
\par Considering the fact that the present universe might have been
formed out of a system of ficticious self-gravitating particles,
fermionic in nature, each of mass $m$, we are able to obtain a
compact expression for the radius $R_0$ of the universe by using a model
density distribution $\rho (r)$ for the particles which is singular
at the origin. This singularity in $\rho (r)$ can be considered to be
consistent with the socalled Big Bang theory of the universe. By
assuming that Mach's principle holds good in the evolution of the
universe, we determine the number of particles, $N$,
of the universe and its $R_0$, which are obtained in terms of the mass $m$
of the constituent particles and the Universal Gravitational constant
$G$ only. It is seen that for a mass of the constituent particles
$m\simeq 1.07\times 10^{-35} g$ the age of the present
universe,$\tau_0$, becomes $\tau_0 \simeq 20\times 10^9 yr$, or
equivalently $R_0 \simeq 1.9\times 10^{28} cm $. For this
$m$, the total number of particles costituting the present universe is
found to be $N \simeq 2.4 \times 10^{91}$ and its total mass
$(M \simeq 1.27916 \times 10^{23} M_{\odot})$, $M_{\odot}$
being the solar mass. All these numbers seem to be quantitatively
agreeing with those evaluated from other theories. Using the present
theory, we have also made an
estimation of the variation of the universal gravitational constant
$G$ with time which gives $({\dot G \over G}) =-9.6\times 10^{-11}
yr^{-1}$. This is again in extremely good agreement with the results
of some of
the most recent calculations. Lastly, a plausible explanation for the
Dark Matter present in today's universe is given. Assuming neutrinoes
to be one of the most possible candidate for the Dark Matter, we
have estimated the ratio of the number of neutrinos to nucleons and
the number of neutrinos per unit volume of the
universe, which respectively gives $({N_{\nu} \over N_n})\sim
3\times 10^9$, $({N_{\nu}\over V})\sim 500$. Both these numbers seem
to be in agreement with
the findings of the recent observations. The present calculation
gives a mass for the neutrino to be $m_{\nu}\sim 1.7 \times
10^{-32} g $ or equivalently, $8eV$, which is the right order of
magnitude, as speculated by several workers.
\eject

\noindent {\bf I. Introduction}

The universe, as we know till today, consists of as many as $10^{11}$ galaxies. These 
galaxies are of course complicated structures, bound gravitationaly, each consisting of upto 
$10^{11}$ stars as well as gas clouds. Each star is a nuclear power-house, and our sun is a 
typical star, situated about halfway out towards the edge of the disc of our galaxy called
the Milky way. The Earth is one of the few planets bound to the sun through
gravitational forces. There may well be other form of matter in the universe besides the
galaxies we see. For instance, the galaxies that have ceased to radiate, black holes
of all sizes and intergalactic dust and gas. However, firm experimental evidence for these is 
lacking.

An understanding of the universe has not only remained as a fascinating problem
for scientists since a longtime, but also its birth has been a challenging problem
for them. There is no satisfactory theory yet relating to the evolution of the
universe. The evolution of the universe from nothing is what is known
as the Big Bang
Theory [1]. According to this theory, the universe started with a huge explosion
from a superdense and a superhot stage. Mathematically, a superdense
state corresponds to imagining of a singularity in the density
distribution function at the origin of the co-ordinate system. 
The very concept of having an infinite density
at the instant of the Big Bang
is not to be taken too seriously. Because, as one looks backward
in time with density going up and up, one is led far away from the condition under
which the laws of physics, as we know them, were developed. Thus, it is quite likely
that at some point, these laws become totally invalid and it is a matter of guess
work to discuss what happened at still earlier times. Nonetheless, as the history of the universe is 
extrapolated, the density increases without limit for as long as the known laws of physics
apply. Indeed, most cosmologists today are reasonably confident about our understanding
of the history of the universe back to one micro second $(10^{-6} sec)$ after the Big Bang.
The goal of the cosmological research involving Grand Unified Theories (GUTs) [2] is to
solidify our understanding back to $10^{-35}$ secs after the Big Bang. Having
discussed the key features of the Big Bang, it is natural to ask now what evidense can be 
found to support it. Before going to discuss about this, one might mention here about
the expansion of the physical universe which is being considered to be one of the greatest
discoveries of the century. The concept of an expanding universe was gradually
found to be accepted by means of a famous law known as Hubble's law [3] which states that each
distant galaxy is receeding from us with a velocity $v$ which, to a high
degree of accuracy, is proportional to its distance $d$. That is,
$$v = H_0d, \eqno{(1)}$$
where $H_0$ is known as the Hubble contant, which is considered to be a constant
presumembly because, it remains approximately constant over the lifetime of an
astronomer. However, value of $H_0$ changes as the universe evolves. Experimental evidence
in favour of the expansion of the universe follows from the fact that light rays received
from distant galaxies are red shifted. There is, ofcourse, a great uncertainty in the 
measurement of the value of $H_0$. The inaccuracy is mostly due to the great difficulties
in calculating the distances to the galaxies. However, as regards the redshifts are 
concerned, they can be determined very accurately from the Doppler shift of the 
spectral lines of the light rays coming from distant galaxies. An accurate measurement
of $H_0$ has important consequences. In the context of Big Bang model, an erroneously high
value for $H_0$ (the expansion rate) implies an erroneously low value for the 
age of the Universe. Leaving aside the fact that $H_0$ is not very accurately known,
there are certain compelling reasons to suggest that the age of the universe
should be around 20 billon years [2]. There is also an uncertainty in accurately measuring
the mass density of the universe, which is considered to be an important quantity
in calculating the history of the universe. Because, it determines how fast does the
cosmic expansion slow down under the influence of gravity. Having
said so much regarding the expansion of the Universe, we are going to
cite some of the evidences in support of the Big Bang theory.
There are two significant observational
evidences in favour of the Big Bang theory. The first one is the observation of the
cosmic background radiation. The first observation of the cosmic background
radiation in the microwave region by Penzias and Wilson [4] was the most important
cosmological discovery since Hubble established the expansion of the universe.
As we all know, all hot matter emits a glow, just like a glow of hot coals
in a fire. Thus, an early universe would have been permiated by a glow of light emitted
by the hot matter. As the universe expanded, the light would have
redshifted. Besides, it also got cooled. Today,
the universe would still be bathed by the radiation, a remnant  of the intense heat of the
Big Bang, now redshifted into the microwave part of the spectrum. This prediction
was confirmed when Penzias and Wilson [4] discovered a background of microwave
radiation with an effective temperature of $\approx \ 2.7^0K$. The characteristic
of this radiation is that it is almost absolutely isotropic, meaning there by that this
radiation is not due to stars or galaxies. The only plausible explanation for
the origin of the radiation is that the universe had perhaps passed through a state of
very high density and high temperature in its early stage. The second important
piece of evidence supporting the Big Bang theory is related to the calculation of what
is called Big Bang nucleo-synthesis, which deals with the calculation of the rates of
different nuclear reactions that took place in the early stage of the universe. Since
it is assumed that the early universe was very hot, it is not possible to have
stable nuclei formed at that stage. At about 2 minutes after the Big Bang, there
were virtually no nuclei at all, because the temperature of the universe
was then very high lying between $(10^9\sim 10^{10})$K. At that state, the universe
was filled with hot gas of photons, neutrinos with a much smaller
density of protons, neutrons and electrons. As the universe
cooled, protons and neutrons began to coalesce to form nuclei. From the nuclear
reaction rates, one can calculate the expected abundances of the different
kinds of nuclei that would have been formed. One finds that most of the
matter in the universe would remain in the form of hydrogen. About $25\%$ of the mass of the
matter would have been converted to helium and trace amount of other nuclei would
also have been produced. Most of the types of nuclei that we observe in the
universe today were produced much later in the history of the universe, in the
interior of stars and in supernova explosions. Hence, the lightest nuclei were produced
primarily in the Big Bang and it is possible to compare the
calculated abundances of these with direct
observations. These calculations, depend on the density of protons and neutrons in the
universe, a quantity which can be estimated only roughly by astronomical observations. It is
interesting to see that there is a broad range of values for the density of protons and neutrons
for which the calculated values for the abundances of light nuclei are in excellent
agreement with the observed values [2]. If there was never a Big Bang, there would
be no reason whatsoever to expect that helium-4 would be $10^8$ times
as abundant as lithium-7. It might just as well have been the other
way around. But, when calculated in the context of the Big Bang
theory, the ratio works out just right.

So far we have been largely concerned with the observational informations about
the universe such as the expansion of the universe, the cosmic background radiation
and the primordial abundance of the elements. We now face the obvious question
like whether the universe will go on expanding forever or will it eventually fall back
onto itself ? The answer to this question that is, the eventual fate of the
universe depends on how much matter it contains on its present size and how fast it is
expanding. The more matter there is, the greater are the gravitational forces holding
the universe together and more likely it is that the universe is finite and will
eventually collapse back onto itself. The greater the present size of the universe,
the farther are the galaxies and the rest of the matter are separated from each
other, and the weaker is the gravitational attraction. This gives rise to an increase
of the likelyhood that the universe is infinite and will expand forever.
Finally, the greater the velocity of expansion, the harder it is for gravity
to slow the expansion to zero and reverse it. Again this contributes towards
the universe being infinite and ever expanding. Thus, one has to determine the relative
contributions of each of the three factors such as the cosmic mass, size and expansion velocity
to the evolution of the universe, and then to use them to decide, on theoretical
grounds, what the density of the universe will be.

The best value of the Hubble constant known at present is $H_0\approx
100 h_0 kms^{-1} Mpc^{-1}$, $0.5 \leq h_0 \leq 1$
[5]. For this rate of expansion, Einstein's General Theory of Relativity predicts that the
universe is finite and closed [6] if the average density of matter $\rho_0$ in the
Universe is greater than $5\times 10^{-30} g/cm^3$, it is infinite and open if
the average density is less than this value; and it is infinite and flat if the
average density is equal to it, where the quantity $5\times 10^{-30} g/cm^3$ is
called the critical density, $\rho_c$, of the universe and this is related to the
Hubble constant $H_0$ as $\rho_c = \bigg ({3 H_0^2\over 4\pi G}\bigg
) q$, [1],where $q$ is the decceleration parameter. The
measurement of $\rho_0$ is an extremely difficult task, because the universe
consists of all sorts of matter and all of them contribute to $\rho_0$. 
Considering the fact that stars form into galaxies, galaxies into clusters, and
clusters into rich super clusters, the average density of matter in the universe
is to be obtained by adding up the masses of galaxies of superclusters in our vicinity
of the universe and dividing it by the corresponding volume. In other words, for
some fixed regions of the sky, one counts the number of galaxies and from that
the galaxies per volume, $n_G$, is calculated. Determination of the
mean mass of a galaxy,
$M_G$, is done following the method as used for the sun. For example,
it is known that the
mass of the sun is measured by using the properties associated with the motion
of the planets around the sun. Like the planets in the solar system, the stars in many galaxies revolve 
around the centre of the galaxy. Hence, once we know the rotating  velocity
of some of the stars and their distance from the galactic centre, the mean
mass of a galaxy, $M_G$, can be determined. Here it is assumed that
the mass of a galaxy is concentrated at its centre. The average density of matter in the
universe
determined by just adding the masses of all the galaxies in a large volume
of space of the sky is found to be $\approx 3 \times
10^{-31} g/cm^3$ [7]. This is, what is called the density of the
observed (luminous) matter
in the universe, which is about 20 times too low to account for a closed universe.

The above value for the average density of the universe has been obtained
by assuming that cosmic matter is mainly concentrated in galaxies. This
need not be the case, because, the space between galaxies is not a vacuum,
but contains gas or extinguished stars. Looking at Zwicky's work [8],
one also finds that the measurement of the mass of the cluster of
galaxies can be done in two ways. One way is via, the luminosity and the other
method is based on dynamics, that is , by measuring the relative
velocities among the galaxies. Since there is a definite relation
between the luminosity of a galaxy and its mass, one can deduce the
mass of the galaxy from luminosity measurements. Then by adding the
masses of the member galaxies, one obtains the total mass of the
cluster. In the method based on measuring the relative velocities
among the galaxies, the mean relative velocity of the whole cluster
is determined by its mass. Turning the problem around, one can obtain
the mass of the cluster from the velocity measurements.
It has been seen that the 
mass found by these two methods differ greatly. For example, the dynamical
masses of the coma cluster [8] is found to be 400 times the luminosity mass.
This result can be interpreted by the fact that the main mass of the coma cluster is
not contributed by the visible galaxies, but by a large amount of invisible
matter within the cluster. The mass measured from the luminosities includes the
mass in the light emitting regions and does not include the mass of any matter
that exists in regions not emitting light. This is what has been identified as the Dark Matter or the
Missing Matter. The question is, just how much Dark Matter is there in the universe
and this, when added to the visible mass, can it give rise to an average density
exceeding the critical density $\rho_c$ ? That would decide whether the universe should be
finite, closed and eventually collapsing. However, at present, the estimate of
the average density of the universe is too uncertain to allow to draw a 
definite conclusion. Today, it is mostly accepted that the mass of the
universe is mainly contributed by Dark Matter and more than 90 percent of the
matter in the universe is invisible. There is some evidence [9] and it has been
indeed the current fashion to believe that the Dark Matter is not the ordinary
matter made of protons, neutrons and electrons, but rather some new stable
exotic neutral elementary particles. Although we know a great deal about the
behaviour of the dark matter we have no observational clues as to what it may be. Considerable
ingenuity is being expended now on devices which will detect the Dark Matter
directly if the particles interact with ordinary matter through weak nuclear force, as is expected for 
some candidates. In order to say something about the most probable candidate for the Dark Matter
we introduce a parameter $\Omega$ which is defined as the ratio of
current density $\rho_0$ of
the Universe to the critical density $\rho_c$ ,$\Omega = (\rho_0/\rho_c)$. If $\Omega < 0.2$,
we imagine of a universe in which the Dark Matter is of the normal kind, that is the types
that make up galaxies, star and atoms etc. This is what is termed as baryonic.
The limits imposed by nucleogenesis[2] tells that the Dark Matter cannot
be baryonic. Also if $\Omega \approx 1$, then the Dark Matter may not be
baryonic. Non-baryonic forms of matter that have been suggested include
neutrinos, gravitinos, photinos, sneutrinos, axions, and magnetic
monopoles etc. Out of
these, neutrinos are considered to be one of the first appealing candidates for the
Dark Matter. Neutrinos are known to exist and they are usually assumed to have
zero rest mass. The universe is assumed to contain a vast number of neutrinos. If a neutrino has a tiny
rest mass, it can far exceed the mass of the baryonic component, and
thus, becoming the
dominant component of the cosmic mass. It is the rest mass of the neutrinos, however small that
may be, that makes the universe finite and closed. A recent study by Schramm and Steigman [10] seems
to suggest that the mass for the neutrino could be between 4 to 20 $eV/c^2$. Further, if
neutrinos have non-zero rest masses, then it is possible that there are neutrino oscillations 
[1] between the various types of neutrinos. These neutrinos do not participate
in electromagnetic interactions. Since they interact very weakly with
matter, they cannot
be detected in today's laboratory experiments. Such features
match with the properties expected
for Dark Matter.

Considering the fact that the major constituents of the present universe are
the neutrinos, which might have been formed out of some ficticious particles
of certain mass that had filled the early universe at a time some two minutes after the Big Bang,
we, in this paper, have tried to calculate the mass, mean density and radius of the
universe by treating the universe just as a system made out of these
particles which are self gravitating.. The present calculation is based on making an intuitive
choice for the distribution of particles in the universe,
characterized by a distribution
function having a singularity at the origin. Such a form of single
particle density seems to be consistent
with the concept like the Big Bang theory of the Universe. The present calculation
is the result of our earlier study [11] where such a form of singular
density distribution has been used to calculate
the binding energy of a system of self gravitating
particles like the neutron stars and white dwarfs etc. having no
source for nuclear 
power generation present at their cores. In all these cases our theory has
proved to be of a great success
due to the fact that it has correctly reproduced the results for the binding
energies and the radii of the neutron stars etc, that agree with
those obtained by other workers. In this work, we have also been able to make
an estimate of the critical mass of a neutron star beyond
which the black hole formation should take place. 
In the present work, we assume the ficticious particles to be fermionic
in nature, carrying a tiny mass $m$, which interact among themselves through
gravitational forces. Considering the fact that the present age of the universe is about
20 billion years [2], we have been able to make an estimate
for the radius of the universe by adjusting the mass of these ficticious
particles. The age of the Universeis determined using the fact that the universe has been
expanding with a speed roughly equal to the velocity of light $'c'$. With this, we
not only arrive of a value of $10^{28}$ cm for the radius of the universe
but also we obtain a value of its mass of about $10^{23} M_{\odot}, \ M_{\odot}$ being
the solar mass $(M_{\odot} = 2\times 10^{33} g)$. All these numbers
seem to be
matching very nicely with the corresponding results known from other theories. Our
estimated result for the average mass density of the present universe comes out to be
$\approx 10^{-29} g/cm^3$, which is supposed to be the case as par expectation.
In Sec.II of the paper the mathematical formalism for the present calculation
is presented. In Sec.III all the results of the present theory have been
derived. In Sec.IV the implication of having a large value for the average
density of the universe is being discussed. Our calculated value for the mass of the
particles responsible for the formation of the Dark Matter seems to
be agreeing with the mass of neutrinos as speculated by several other
workers [1]. In this
section, we have also presented our estimated values for
the ratio of the variation of the universal gravitational constant $G$ with
time to $G$ itself, that is $({^.G \over G} )$. This is again found to be in extremely good agreement
with those obtained from some of the most recent calculations [5].

\vskip .4 true cm
\noindent {\bf II. Mathematical Derivation for the Radius of the Universe}

If we visualize the present universe to be a system composed of some self gravitating
ficticious particle, each of mass $m$, interacting through pair-wise gravitational
interactions, then the Hamiltonian of the system can be written as
$$H = - \sum^N_{i=1}({\hbar^2\over 2m})\nabla_i^2+{1\over 2}\sum^N_{i=1}
\sum^N_{j=1} v (\mid\vec X_i-\vec X_j\mid),\eqno{(1)}$$
where i is not equal to j and 
$v(\mid\vec X_i-\vec X_j\mid) = - g^2/\mid\vec X_i-\vec X_j\mid$ ,having $g^2=Gm^2$
,$G$ being Newton's universal gravitational constant. The assumption that goes in while
writing (1) is that there is no radiation source present at the centre of
the universe. The particles which constitute the universe are some
kind of fermions whose mass is
estimated by using certain known results that are considered to be approximately
true for the universe. As the universe evolves with time, it is these ficticious particles
that might have, latter on, given rise to the production of heavy particles found
in the present universe and to those known to constitute
the the Dark Matter contained within it.
In the Thomas-Fermi approximation [12], we evaluate the total kinetic energy of the system
using the relation :
$$<KE> = \bigg ({3\hbar^2\over 10m}\bigg )(3 \pi^2)^{2/3}\int d\vec X [\rho(\vec X)]^{5/3},\eqno{(2)}$$
where $\rho (\vec X)$ is the single particle distribution function
which the constitutent particle obey within the universe, such that 
$$\int d\vec X \rho(\vec X) = N, \eqno{(3)}$$
$N$ being the total number of these particles in the universe. The total potential energy
of the system in the Hartree-approximation is obtained as
$$<PE> = - ({g^2\over 2})\int d\vec X d\vec X'{1\over \mid\vec
X-\vec X' \mid} \rho(\vec X)\rho(\vec X')\eqno{(4)}$$
Evaluation of the integrals shown in (2) and (4) is done using a single
particle density distribution of the form:
$$\rho(\vec X) = {{A e^{-x}}\over x^3},\eqno{(5)}$$
where $x = (r/\lambda)^{1/2}, \ r = \mid\vec X\mid$. Such a form of the single particle density for the
ground state has been used by us before [11] while calculating the binding energy of a
neutron star and its radius. Latter on [13], it was used to obtain a quantum mechanical
derivation for the Schwarzschild radius of a blackhole, thereby
indicating further justification of its use. After evaluating the integrals given in (2)
and (4), the expression
for total energy of the universe, $E_0 (\lambda)$, becomes
$$E_0(\lambda) = ({C\over\lambda^2}) - ({D\over\lambda}),\eqno{(6)}$$
where
$$ \eqalign{C & = ({12\over 25\pi}) ({\hbar^2\over m}) ({3\pi N\over 16})^{5/3},\cr
D & = ({g^2N^2\over 16})\cr}\eqno{(7)}$$
Now, minimising $E_0 (\lambda)$ with respect the $\lambda$ and then
evaluating it at
$\lambda = \lambda_0$, where $\lambda_0$ is the value of $\lambda$ at which
the minimum occurs, the total binding energy of the universe, corresponding to its
lowest energy state, becomes
$$E_0 = - (0.015442) N^{7/3} ({mg^4\over \hbar^2})\eqno{(8)}$$
In view of our earlier work [13] concerning the
derivation of the Schwarzschild radius, here also we 
identify $2\lambda_0$ result with the radius $R_0$ of the universe, whose expression
becomes
$$R_0 =2\lambda_0 = ({\hbar^2\over mg^2}) \times (4.047528)/N^{1/3}\eqno{(9)}$$

\vskip .5 true cm
\noindent {\bf III. Results of the Present Theory}

We now try to recall the Mach's principle [14], according to which all
inertial forces are due to the distribution of matter in the Universe.
This can be conveniently put forth in a mathematical form through the following relation:
$$ {GM\over R_0 c^2} \approx  1\eqno{(10)}$$
It has been categorically stated by Collins and Hawking [15] that Mach's relation appears
to hold good in the evolution of the universe. Furthermore, within
the frame work of the expanding
universe, it also suggests that $G$ cannot be constant in time. Using (10) in
(9), one arrives at
$$N = 2.8535954 \bigg ({\hbar c\over Gm^2}\bigg )^{3/2}\eqno{(11)}$$
Substituting the (11) in (9), the expression for $R_0$ becomes
$$R_0 = 2.8535954 \bigg ( {\hbar\over mc}\bigg ) \bigg ({\hbar c\over Gm^2}\bigg )^{1/2}\eqno{(12)}$$
Using (12), we have made an estimation for the the radius of the
prsent 
universe, $R_0$, by varying $m$, the mass of the ficticious particle
and using the measured value for the gravitational constant
$G$, that is by taking
$G = 6.67 \times 10^{-8}$ $dyn \ cm^2 \ g^{-2}$. We have chosen a set of four values for
$m$ and calculate $R_0$, from which the age of
the Universe $\tau_0$ is estimated using the relation $R_0 \simeq c\tau_0$. All
these are shown in Table-I of this paper for which $\tau_0 $ has been
considered to 20, 15, 10 and 5 billion years. 
For these $m$ values, the number of particles $N$ in the universe is calculated with the help
of (11), which enables us to determine the total mass of the universe, $M_0$. These
are also given in Table-I. Using the results for $M_0$ and $R_0$, the average mass
density of the present universe is estimated. This is shown in Table-II of this paper.

We now try to calculate the variation of the gravitational constant $G$ with respect to
time $\dot {G(t)}$. For that, we, use the expression for $G$, as
found from (12), which is obtained as
$$G = K/R_0^2,\eqno{(13)}$$
where
$$K = (8.1430067) ({\hbar^3\over m^4c})$$
Further ,we assume that the above expression for $G$ is also valid for
anytime 't' provided we take the value of $R_0$ at that time.
Using (13), we therefore, find 
$$\dot G = \bigg ( {\partial G\over\partial R_0}\bigg ) \times \bigg ({\partial R_0\over
\partial t}\bigg ) \simeq c \bigg ({\partial G\over\partial R_0}\bigg )= {-2cK\over
R^3_0}\eqno{(14)}$$
Following this, we make an estimation of $({\dot G\over G})$, which gives
$$\bigg ({\dot G\over G} \bigg ) = - 9.6 \times 10^{-11} yr^{-1}\simeq -1\times
10^{-10} yr^{-1}\eqno{(15)}$$
The above value is in extremely good agreement with one of the recent
estimates [1]
and is also consistent with that of P.M.Muller and Hoyle and Narlikar [5]. This is
roughly also the value reported by Van
Flandern [16], following an analysis of the data relating to the effects of
tidal friction on the elements and shape of the lunar
orbit of the earth-moon system, within the frame work of the Dirac cosmology.
However Shapiro et al [17], have arrived at a limiting
value for $\mid({\dot G\over G})\mid \leq 4\times 10^{-10} yr^{-1}$, based
on a method which is used
for monitoring the planets for a possible secular increase in their orbital
periods by employing a radar reflection system between the
Earth,Venus and Mercury.
This very result obtained by Shapiro et al, is found to be almost ten times greater than the
value predicted by Dicke from the Brans-Dicke scalar tensor theory
[18]. This means that the Brans-Dicke theory gives a value for $\mid
({\dot G\over G})\mid \approx 4\times 10^{-11} yr^{-1}$.In
order to see whether we can at all reproduce some of these results
from our theory, 
we have made an estimation of $({\dot G\over G})$ by varying the
mass $m$ of the
constituent particles. These are shown in Table-II of the paper.
From this, it can be
seen that as $m$ increases, the age of universe, $\tau_0$, decreases.
This amounts to also a further decrease in the value of $({\dot G\over G})$.
Accepting the fact that the age of the universe is around 20 billion
years, one would find that the Brans-Dicke result
would correspond to an age greater than twenty billion years.

\vskip .4 true cm
\noindent {\bf IV. \ Discussion of Results and Conclusion}

In arriving at the various results of the present theory, we have
accepted the value for the age of the universe $\tau_0$ to be $\sim 20\times 10^{9} yr$.
The mass of the constituent
particles corresponding to this is found to be $m\sim 1.07
\times 10^{-35} g$. It is also
to be noted that in the present theory we have treated the
constituent particles as fermions.
Since we have mentioned that the neutrinos could be one of the most probable candidates
for the Dark Matter present in today's universe, it is therefore
expected that the neutrinos are to be formed out of these
ficticious particles whose masses are  $m\simeq 1.07 \times
10^{-35} g$, or equivalently of energy 0.006 eV, instead of zero, as
thought to be the case [7]. For a mass of $m\sim 1.07
\times 10^{-35} g$, the radius of the universe becomes
$R_0 \sim 1.9\times 10^{28}$ cm, which is considered to be an acceptable result,
and total
mass of the universe $M_0$ becomes $M_0 \sim (1.3\times 10^{23}) M_{\odot}$,
$M_{\odot}$ being the solar mass. If we assume the mass $M_0$ to be solely
due to the nucleons that are there in the today's universe, it would correspond to
$\sim 1.5 \times 10^{80}$ nucleons to be present. This is nothing but
the famous Eddington result [19]. 
Using this, the ratio of the number of ficticious particles constituting universe
to the number of equivalent nucleons is found to be $\sim 1.6 \times 10^{11}$.
This is of the right order of magnitude such as $10^{10}$, the value that has been speculated for the number of photons 
(neutrinos) per nucleon in the observed universe [2]. We would like
to mention here that in arriving at the above value we have assumed
that all the ficticious particles involved in the formation of the early universe
have been subsequently converted to nucleons and heavy nuclei etc that are found in the today's
universe in the form of visible matter. Let us now look at our calculated value for the average mass density of
the universe from Table-II corresponding to $m \sim 1.07 \times
10^{-35} g$ which gives $\sim 0.9\times 10^{-29}g/cm^3$. This, we find to be almost
thirty times larger then the observed mass density ,$\rho_0$ ,of the
present universe, $\rho_0 \sim 3\times 10^{-31} g/cm^3$
where the latter is being thought to be arising solely due to nucleons and other
heavy elements etc, present. Let us now assume
that out of the entire mass $M_0$ of the universe, some percentage of it has
gone into the formation of the nucleons and other heavy elements
present today  and the remaining part has been lying in the form
of Dark Matter in the present universe. This amounts to saying that the total number of
ficticious particles responsible in the formation of the early universe have
been subsequently converted to appear in the present universe as
nucleons and other heavy nuclei and some other
form of particles contributing to the Dark Matter of the universe. These
particles could be most probably the neutrinos, as speculated by several workers.
Let us accept the view that not more than three to ten percent of the entire mass of the
universe is contributed by the nucleons and other heavy nuclei [7],
the contribution from the heavy nuclei being negligibly small
compared to that of the nucleons. This would therefore mean that the
remaing  ninety  to ninety-seven percent of its total mass constitutes what is known 
as the Dark Matter of the universe. Having accepted this picture, we have calculated
the ratio of the number of neutrinos to nucleons, $(N_{\nu}/N_n)$, and the
neutrino number per unit volume of the universe, $(N_{\nu}/V)$, etc, $V$ being the
volume of the universe, by varying the total mass of the existing
nucleons so as to lye within
three to ten percent. These calculations have been repeated for different
values of the age of the universe kept within five to twenty billion years.
These are exhibited in  table-III as shown below.

From the above table we find that the calculation done by assuming
that only three percent of the total mass of the universe constitutes the observed
density of the universe, is the most appropriate one. Because, for this case, the average
density of the nucleons (which constitute the observable matter of the universe)
comes out to be $\rho_0\sim 3\times 10^{-31} \ g/cm^3$, which is
considered to be the most accepted value. This corresponds to an age 
of twenty billion years for the universe, and for this the total nucleon number of the universe
becomes $\sim 4.6 \times 10^{78}$ ,which again nicely agrees with the most speculated value [14].
This being so, the remaining ninety seven percent of the total mass of the universe is 
considered to be solely due to the Dark Matter present within it. Accepting the
fact that this mass is generated by the neutrinos present in the universe,
we have made an estimation of the ratio $(N_{\nu}/N_n)$ and $(N_{\nu}/V)$ by
varying the mass of the neutrinos. From table-III, one can see that
for a neutrino mass of
$m_{\nu} \sim 1.73 \times 10^{-32}$ g, one arrives at $(N_{\nu}/N_n) \simeq 3.1 \times
10^9$ and $({N_{\nu}\over V})\simeq 500$. Both these numbers show fantastic agreement
with the findings of the recent observation [7]. A mass of $m_{\nu} = 1.73 \times
10^{-32} g$ corresponds to an energy of $8eV$, which is again found to be the right order of 
magnitude for the neutrino mass as speculated by recent measurements by several workers.

Considering the value for the average mass density of the present universe
(shown in Table-II) which comes out to be $\rho_U \simeq 9 \times
10^{-30} g/cm^3$, one can clearly see that
it is obviously much larger than the observed mass density $\rho_0$
$(\rho_0\simeq 3\times 10^{-31}g/cm^3)$. Since this also  
exceeds the critical density $\rho_c \ (\rho_c \sim 5 \times 10^{-30} \ g/cm^3$)
we would expect the present universe to be close one. Of course,
the results of this calculation should not be taken up conclusively,
since a thorough analysis of
this is needed to make a definite statement. A rough estimate of the Hubble
constant $H_0$ from the present theory $(H \approx {1\over \tau_0})$
gives a value of $48 km/sec/Mpc$ [Table-II]. This
is also in good agreement within the range of speculated value
for $H_0$. As we have seen,a value of $48 km/sec/Mpc$ for $H_0$
corresponds to an age $20\times 10^9 $ yrs for the present universe,
and a $\rho_c \approx \ 5\times 10^{-30} \ g/cm^3$ $[\rho_c = ({3
H_0^2 \over {8\pi G}})]$. Since the present theory gives rise to $\approx 500$ neutrinos
per $cm^3$, this is to be equal to the number of photons per $cm^3$, in view
of the common belief that there are an equal number of photons and neutrinos in the
present universe [3]. 

To conclude, we have shown that the results of the present
calculation are obtained by choosing
a singular form of single particle density for the particles
constituting the universe and a singular
density is consistent with the idea relating to the Big Bang theory. Since the
standard Big Bang theory so far has not succeeded to explain for the
ratio of $\approx 10^9$ for (number of photons/number of nuclei) [2] and here 
we do reproduce the above number correctly, the present
work seems to be justifying the so called Big Bang theory of the universe.

Although the General Theory of Relativity assumes that $G$ has to remain constant, there seems
to be some compelling reason to think that $G$ must vary with time, hence with the age
of the universe. From the present theory we find that $G \alpha {1\over \tau^2}$ 
(vide Eq.(13)), $\tau$ being the age of the universe, whereas it is usually
assumed that $G$ is inversely proportional to $\tau$ [1]. It would be
therefore interesting to see what kind of new features it can exhibit
if $G$ or the laws of nature change with time.

In making an estimation of the Hubble constant $H_0$ we have approximated the velocity
of expansion of the universe to be equal to the velocity of light `c'. Since
after accounting for the presence of the  Dark Matter in the universe, the actual particle density
of the universe might exceed the critical density $\rho_c$, one would
expect the present universe to show a contraction after certain
stage. A contraction of the Universe would mean a gradual decrease in
the velocity of expansion. It is therefore necessary to have an
estimate of the deceleration parameter $q$ associated with the contraction. For that
the present theory needs to be carefully studied.
\vfill
\eject
\noindent {\bf References}
\item {[1]} G. Contopoulos and D. Kotsakis, {\it Cosmology},
(Springer-Verlag,Heidelberg, 1987)
\item {[2]} Alan  H. Guth, in {\it Bubbles,voids and bumps in time: the new
cosmology} ed. James Cornell (Cambridge University Press,Cambridge,1989)
\item {[3]} Fang Li Zhi and Li Shu Xian ,{\it Creation of the
Universe},(World Scientific,Singapur,1989)
\item {[4]} A.A.Penzias and R.W. Wilson,{\it Astrophys.Jour.} {\bf 142}, (1965),419
\item {[5]} J.V.Narlikar,{\it Introduction To Cosmology}, (Cambridge
University Press,London,1993)
\item {[6]} G.S.Kutter ,{\it The universe and life},(Jones and Bartlett,USA,1987)
\item {[7]} I.Novikov,{\it Black holes and the Universe},(Cambridge
University Press,Cambridge,1990)
\item {[8]} F.Zwicky,{\it Helv.Phys.Act},{\bf 6},(1933),10
\item {[9]} James E.Gunn in  {\it Bubbles,voids and bumps in time: the new
cosmology} ed. James Cornel (Cambridge University Press,Cambridge, 1989)
\item {[10]} D.N.Schramm, {\it Phys.Today} ,{\bf 36}, (1983), 27.
G.Steigman,{\it Ann. Rev. Astron. Astrophys.} ,{\bf14}, (1976), 339
\item {[11]} D.N.Tripathy and Subodha Mishra,Astr-Ph/9612097
\item {[12]} L.D.Landau and E.M.Lifshitz,{\it Quantum
Mecanics},(Pergamon Press,Oxford,1965)
\item {[13]} D.N.Tripathy and Subodha Mishra ,to be published.
\item {[14]} P.S.Wesson, {\it Cosmology and Geophysics},(Adam Hilger Ltd,Bristol,1978.
\item {[15]} C.B.Collins and S.W.Hawking, {\it Astrophys.J.} {\bf
180} (1973),317.
\item {[16]} T.C.Van Flandern,{\it Mon. Not.R.Astron. Soc.}, {\bf 170},(1975),333.
\item {[17]} I.I.Shapiro,{\it BAAS},{\bf 8},(1976),308.
\item {[18]} C.Brans and R.H.Dicke,{\it Phys.Rev},{\bf 124},(1961),125
\item {[19]} E.R.Harrison,{\it Cosmology}, (Cambridge: Cambridge University Press,1981)
\vfill
\eject

 \magnification=\magstep1
\hsize= 23.0 true cm
\vsize= 18.0 true cm
\centerline {\bf {TABLE - I }}
\midinsert$$\vbox{\offinterlineskip
\halign{&\vrule#&\strut\ #\ \cr
\noalign{\hrule}
&\hfil $m\times 10^{-35} g$  \hfil&&\hfil $R\times 10^{28} cm$   \hfil&&\hfil $N\times 10^{91}$
\hfil&&\hfil $M\times 10^{56} gm$  \hfil&&\hfil $\tau_0\times 10^9 yr$    &\hfil\cr 
height3pt&\omit&&\omit&&\omit&&\omit&&\omit&\cr
\noalign{\hrule}
&\hfil 1.07299  \hfil&&\hfil 1.896  \hfil&&\hfil 2.38429   \hfil&&\hfil  2.55832 \hfil&&\hfil  20   &\hfil\cr 
height3pt&\omit&&\omit&&\omit&&\omit&&\omit&\cr
&\hfil 1.23891  \hfil&&\hfil 1.422  \hfil&&\hfil 1.54865   \hfil&&\hfil 1.91875 \hfil&&\hfil 15   &\hfil\cr 
height3pt&\omit&&\omit&&\omit&&\omit&&\omit&\cr
&\hfil 1.51744  \hfil&&\hfil 0.948   \hfil&&\hfil 0.84297    \hfil&&\hfil 1.27916 \hfil&&\hfil 10   &\hfil\cr 
height3pt&\omit&&\omit&&\omit&&\omit&&\omit&\cr
&\hfil 2.14598  \hfil&&\hfil 0.474  \hfil&&\hfil 0.29804    \hfil&&\hfil 0.639588 \hfil&&\hfil  5  &\hfil\cr 
height3pt&\omit&&\omit&&\omit&&\omit&&\omit&\cr
\noalign{\hrule}
\noalign{\hrule}\noalign{\smallskip}}}$$\endinsert
\vfill
\eject
\hsize= 23.0 true cm
\vsize= 18.0 true cm
\centerline {\bf {TABLE - II }}
\midinsert$$\vbox{\offinterlineskip
\halign{&\vrule#&\strut\ #\ \cr
\noalign{\hrule}
&\hfil $M\times 10^{56} g$  \hfil&&\hfil $R\times 10^{28} cm$
\hfil&&\hfil $\rho\times 10^{-29} g/cm^3$  \hfil&&\hfil $\dot G\times
10^{-25} cm^3 g^{-1} s^{-2}$    \hfil&&\hfil $(\dot G/G) yr^{-1}$ \hfil&&\hfil $H_0 km/sec/Mpc$   &\hfil\cr 
height3pt&\omit&&\omit&&\omit&&\omit&&\omit&&\omit&\cr
\noalign{\hrule}
&\hfil 2.55832  \hfil&&\hfil 1.896 \hfil&&\hfil 0.89609  \hfil&&\hfil -2.11076   \hfil&&\hfil $-1\times 10^{-10}$ \hfil&&\hfil 48.829   &\hfil\cr 
height3pt&\omit&&\omit&&\omit&&\omit&&\omit&&\omit&\cr
&\hfil 1.91875  \hfil&&\hfil 1.422 \hfil&&\hfil 1.59305  \hfil&&\hfil -2.81437   \hfil&&\hfil $-1.3\times 10^{-10}$ \hfil&&\hfil 65.105   &\hfil\cr 
height3pt&\omit&&\omit&&\omit&&\omit&&\omit&&\omit&\cr
&\hfil 1.27916  \hfil&&\hfil 0.948 \hfil&&\hfil  3.58436 \hfil&&\hfil -4.22149   \hfil&&\hfil $-1.9\times 10^{-10}$ \hfil&&\hfil 97.658   &\hfil\cr 
height3pt&\omit&&\omit&&\omit&&\omit&&\omit&&\omit&\cr
&\hfil 0.63959  \hfil&&\hfil 0.474  \hfil&&\hfil 1.43376   \hfil&&\hfil  -8.44304   \hfil&&\hfil $-4\times  10^{-10}$ \hfil&&\hfil 195.316   &\hfil\cr 
height3pt&\omit&&\omit&&\omit&&\omit&&\omit&&\omit&\cr
\noalign{\hrule}
\noalign{\hrule}\noalign{\smallskip}}}$$\endinsert
\vfill
\eject
\hsize= 23.0 true cm
\vsize= 18.0 true cm
\centerline {\bf {TABLE - III }}
\midinsert$$\vbox{\offinterlineskip
\halign{&\vrule#&\strut\ #\ \cr
\noalign{\hrule}
&\hfil $M_n\times 10^{56} g$  \hfil&&\hfil $N_n\times 10^{78}$  \hfil&&\hfil  
$N_{\nu}/N_n\times 10^9$ \hfil&&\hfil $N_{\nu}\times 10^{88}$  \hfil&&\hfil $N_{\nu}/V$   \hfil&&\hfil $M_{\nu} \times 10^{56} g$ \hfil&&\hfil  $m_{\nu}\times 10^{-32}g$   &\hfil\cr 
height3pt&\omit&&\omit&&\omit&&\omit&&\omit&&\omit&&\omit&\cr
\noalign{\hrule}
&\hfil 0.03 (2.55832)  \hfil&&\hfil 4.5875 \hfil&&\hfil 3.344  \hfil&&\hfil 1.53390  \hfil&&\hfil  535  \hfil&&\hfil 2.48157 \hfil&&\hfil 1.6176   &\hfil\cr 
height3pt&\omit&&\omit&&\omit&&\omit&&\omit&&\omit&&\omit&\cr
&\hfil 0.03 (2.55832)  \hfil&&\hfil 4.5875  \hfil&&\hfil 3.125  \hfil&&\hfil 1.43359  \hfil&&\hfil 500   \hfil&&\hfil 2.48157 \hfil&&\hfil 1.7309   &\hfil\cr 
height3pt&\omit&&\omit&&\omit&&\omit&&\omit&&\omit&&\omit&\cr
&\hfil 0.03 (2.55832)  \hfil&&\hfil  4.5875 \hfil&&\hfil 2.813  \hfil&&\hfil 1.29020   \hfil&&\hfil 450   \hfil&&\hfil 2.48157 \hfil&&\hfil 1.9232   &\hfil\cr 
height3pt&\omit&&\omit&&\omit&&\omit&&\omit&&\omit&&\omit&\cr
&\hfil 0.03 (2.55832)  \hfil&&\hfil 4.5875 \hfil&&\hfil 1.000 \hfil&&\hfil 0.45875   \hfil&&\hfil 160   \hfil&&\hfil 2.48157 \hfil&&\hfil 5.4090  &\hfil\cr 
height3pt&\omit&&\omit&&\omit&&\omit&&\omit&&\omit&&\omit&\cr
\noalign{\hrule}
\noalign{\hrule}\noalign{\smallskip}}}$$\endinsert
\vfill
\eject
\hsize= 23.0 true cm
\vsize= 18.0 true cm
\centerline {\bf {TABLE - IV }}
\midinsert$$\vbox{\offinterlineskip
\halign{&\vrule#&\strut\ #\ \cr
\noalign{\hrule}
&\hfil  $\tau_0 \times 10^9 yr$ \hfil&&\hfil $M_n\times 10^{56}g$  \hfil&&\hfil $N_n\times 10^{78}$  \hfil&&\hfil $N_{\nu}/N_n\times 10^9$ \hfil&&\hfil  $N_{\nu}\times 10^{88}$ \hfil&&\hfil $N_{\nu}/V$   \hfil&&\hfil $M_{\nu}\times 10^{56}g$ \hfil&&\hfil $m_{\nu}\times
 10^{-32}g$   &\hfil\cr 
height3pt&\omit&&\omit&&\omit&&\omit&&\omit&&\omit&&\omit&&\omit&\cr
\noalign{\hrule}
&\hfil 20  \hfil&&\hfil 0.03 (2.55832) \hfil&&\hfil 4.587500 \hfil&&\hfil 3.125  \hfil&&\hfil 1.43359  \hfil&&\hfil  500  \hfil&&\hfil  2.48157\hfil&&\hfil 1.73099   &\hfil\cr 
height3pt&\omit&&\omit&&\omit&&\omit&&\omit&&\omit&&\omit&&\omit&\cr
&\hfil 15  \hfil&&\hfil 0.03 (1.91875) \hfil&&\hfil 3.440675  \hfil&&\hfil  3.125 \hfil&&\hfil  1.07521 \hfil&&\hfil 892   \hfil&&\hfil 1.86119 \hfil&&\hfil 1.73099   &\hfil\cr 
height3pt&\omit&&\omit&&\omit&&\omit&&\omit&&\omit&&\omit&&\omit&\cr
&\hfil 10  \hfil&&\hfil 0.03 (1.27916) \hfil&&\hfil 2.293772 \hfil&&\hfil 3.125  \hfil&&\hfil 0.71680  \hfil&&\hfil 2008   \hfil&&\hfil  1.24079\hfil&&\hfil 1.73099  &\hfil\cr 
height3pt&\omit&&\omit&&\omit&&\omit&&\omit&&\omit&&\omit&&\omit&\cr
&\hfil 5  \hfil&&\hfil 0.03 (0.639598)  \hfil&&\hfil 1.146900 \hfil&&\hfil 3.125  \hfil&&\hfil 0.35841  \hfil&&\hfil 8034   \hfil&&\hfil 0.62040 \hfil&&\hfil 1.73099   &\hfil\cr 
height3pt&\omit&&\omit&&\omit&&\omit&&\omit&&\omit&&\omit&&\omit&\cr
\noalign{\hrule}
\noalign{\hrule}\noalign{\smallskip}}}$$\endinsert
\vfill
\eject
\end